\documentclass[prl,twocolumn,superscriptaddress]{revtex4}
\usepackage{amsmath}
\usepackage{color}
\usepackage{epsfig,graphicx,amssymb,bm}
\usepackage{subfigure}

\begin{document}

\title{Preservation Macroscopic Entanglement of Optomechanical Systems in
non-Markovian Environment}
\author{Jiong Cheng}
\author{Wen-Zhao Zhang}
\author{Ling Zhou}
\thanks{zhlhxn@dlut.edu.cn}
\affiliation{School of Physics and Optoelectronic Technology, Dalian University of
Technology, Dalian 116024, People's Republic of China}
\author{Weiping Zhang}
\affiliation{State Key Laboratory of Precision Spectroscopy, Department of Physics, East
China Normal University, Shanghai 200062, People's Republic of China}
\date{\today }

\begin{abstract}
We investigate dynamics of an optomechanical system under the Non-Markovian
environment. In the weak optomechanical single-photon coupling regime, we
provide an analytical approach fully taking into account the non-Markovian
memory effects. When the cavity-bath coupling strength crosses a certain
threshold, an oscillating memory state for the classical cavity field 
(called bound state) is formed. Due to the existence of the non-decay optical
bound state, a nonequilibrium optomechanical thermal entanglement is
preserved even without external driving laser. Our results provide a
potential usage to generate and protect entanglement via Non-Markovian
environment engineering.
\end{abstract}

\pacs{03.65.Yz,42.50.Wk, 03.65.Ud}
\maketitle

\textit{Introduction.}-The investigation of decoherence and dissipation
process induced by environment is a fundamental issue in quantum physics %
\cite{Breuer2002,Weiss2008,DiVincenzo393,Knill409 46}. Understanding the
dynamics of such nonequilibrium open quantum system is a challenge topic
which provides us the insight into the issue of quantum-classical
transitions. Protecting the quantum property from decoherence is a key
problem in quantum information science, therefore a lot of effort has been
devoted in developing methods for isolating systems from their destructive
environment. Recently, people recognize that properly engineering quantum
noise can counteract decoherence and can even be used in robust quantum
state generation \cite{cirac,Eisert}. Meanwhile the features of the
non-Markovian quantum process have sparked a great interest in both
theoretical and experimental studies \cite{Chru104 070406,Xu104 100502, Liu7
931,Chin109 233601,Deffner111 010402}. Numerous quantitative measures have
been proposed to quantify non-Markovianity \cite{Rivas105 050403, Breuer103
210401,Vasile84 052118,Lorenzo88 020102,Chru112 120404}.

As a promising candidate for the exploration of quantum mechanical features
at mesoscopic and even macroscopic scales and for quantum information
procession, cavity optomechanical systems come as a well-developed tool and
have received a lot of attentions \cite{Vitali98 030405,Kippenberg321
1172,Groblacher460 724,Connell464 697,Thompson452 900}. In the theoretical
research of the cavity optomechanical system, the environment is often
treated as a collective non-interacting harmonic oscillators, and the
quantum Langevin equations \cite{Giovannetti63 023812} are developed to
describe the radiation-pressure dynamic backaction phenomena. Significant
progresses have been made in this framework \cite{Vitali98 030405,Genes77
033804,Agarwal81 041803,Rabl107 1}. Almost all of these studies are
focussing on the scenario of memoryless environment. However in many
situations for optical microcavity system, the backaction of the environment
and the memory effect of the bath play a significant role in the decoherence
dynamics \cite{Bayindir84 2140,Hartmann2 849}. Quite recently, a nonorthodox
decoherence phenomenon of the mechanical resonator is also observed in
experiment \cite{Groblacher6 7606}, which clearly reveals the non-Markovian
nature of the dynamics. Therefore, it is the time to investigate the
non-Markovian environment engineering for the nonlinear cavity
optomechanical system so that we can use the memory effects to produce and
protect coherence within it.

In this Letter, we investigate the cavity optomechanical dynamics under
Non-Markovian environment and put forward a method to solve the exact
Heisenberg-Langevin equations where the non-local time-correlation of the
environment is included. We find that when the cavity-bath coupling strength
crosses a certain threshold, the optical bound state is formed, giving rise
to the nonequilibrium dynamics of the entanglement. This remarkable result
indicates the possibility of long-time protection of macroscopic
entanglement via structured reservoirs.

\begin{figure}[tp]
\includegraphics[width=8.25cm]{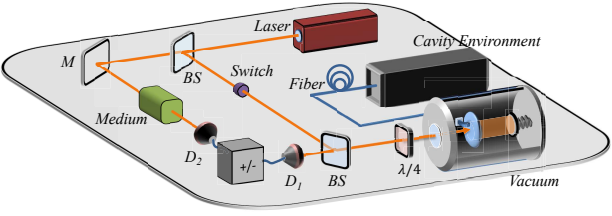}
\caption{(Color online) Example implementations of optomechanical system
setup for study the non-Markovian dynamics: a laser is split into a signal
beam and a local oscillator. The signal (input field) is controlled by the
\emph{Switch} and the output field can be read through photon detectors. The
optomechanical cavity environment is introduced by a low-loss fiber. The
optomechanical cavity is kept at a pressure of $<10^{-3}$ mbar to avoid
residual-gas damping of the mechanical motion.}
\label{fsys}
\end{figure}

\textit{Model.}-We consider a generic cavity optomechanical system
consisting of a Fabry-P\'{e}rot cavity with a movable mirror at one side.
The cavity has equilibrium length $L$, while the movable mirror has
effective mass $m$. As shown in Fig. \ref{fsys}, cavity environment could be
a coupled-resonator optical waveguide which possesses strong non-Markovian
effects \cite{Wu18 18407}, and the micro-mechanical resonator and its
environment could be the device of a high-reflectivity Bragg mirror
fixed in the centre of a doubly clamped $\mathrm{Si_{3}N_{4}}$ beam in vacuum \cite%
{Groblacher6 7606}. The corresponding Hamiltonian of the system can be
written as \cite{Giovannetti63 023812,Genes77 033804} $\hat{H}_{S}=\hbar
\omega _{c}\hat{a}^{\dag }\hat{a}+\frac{\hbar \omega _{m}}{2}(\hat{p}^{2}+%
\hat{q}^{2})-\hbar g_{0}\hat{a}^{\dag }\hat{a}\hat{q}+i\hbar E(e^{-i\omega
_{0}t}\hat{a}^{\dag }-e^{i\omega _{0}t}\hat{a}).$ Here $\omega _{c}$ is the
frequency of the cavity mode with bosonic operators $\hat{a}$ and $\hat{a}%
^{\dag }$ satisfying $[\hat{a},\hat{a}^{\dag }]=1$, while the quadratures $%
\hat{q}$ and $\hat{p}$ ($[\hat{q},\hat{p}]=i$) are associated to the
mechanical mode with frequency $\omega _{m}$. The third term describes the
optomechanical interaction at the single-photon level with coupling
coefficient $g_{0}=(\omega _{c}/L)\sqrt{\hbar /2m\omega _{m}}$. The cavity
is driven by an external laser with the center frequency $\omega _{0}$. The
environment of such system can be described by a collection of independent
harmonic oscillators \cite{Ford37 4419}. The reservoir as well as the
system-reservoir interaction is then given by $\hat{H}_{EI}=\sum_{k}{\hbar
(\omega _{k}\hat{a}_{k}^{\dag }\hat{a}_{k}}+\hbar g_{k}\hat{a}^{\dag }\hat{a}%
_{k}+\hbar g_{k}^{\ast }\hat{a}_{k}^{\dag }\hat{a})+\sum_{l}\frac{\hbar
\omega _{l}}{2}[\hat{q}_{l}^{2}+(\hat{p}_{l}-\gamma _{l}\hat{q})^{2}].$ The
first term is the free energy of the cavity reservoir with the continuous
frequency $\omega _{k}$ as well as the hopping interaction between the
cavity and the environment with the coupling strength $g_{k}$. The second
summation describes a mirror undergoing Brownian motion with the coupling
through the reservoir momentum \cite{Giovannetti63 023812,Ford37 4419}. Here
$\omega _{l}$ is the reservoir energy of the mechanical mode, and $\gamma
_{l}$ stands for the mirror-reservoir coupling.

\textit{Dynamics of the system}.-To achieve a comprehensive understanding of
the decoherence dynamics, one has to rely on precise model calculations. To
this end, by making use of the reference frame rotating at the laser
frequency and tracing out all the environmental degrees of freedom, we can
obtain the Heisenberg-Langevin equations (see Sec. I in the Supplemental
Material)
\begin{subequations}
\begin{eqnarray}
\dot{\hat{a}} &=&-i\Delta _{c}\hat{a}+ig_{0}\hat{a}\hat{q}-\int_{0}^{t}d\tau
f_{c}(t-\tau )\hat{a}(\tau )  \notag \\
&&-i\sum\limits_{k}g_{k}e^{-i\Delta _{k}t}\hat{a}_{k}(0)+E, \\
\dot{\hat{q}} &=&\omega _{m}\hat{p}, \\
\dot{\hat{p}} &=&-\Delta _{m}\hat{q}+g_{0}\hat{a}^{\dag }\hat{a}%
+\int_{0}^{t}d\tau f_{m}(t-\tau )\hat{q}(\tau )  \notag \\
&&+\sum\limits_{l}\omega _{l}\gamma _{l}(\hat{p}_{l}(0)\cos \omega _{l}t-%
\hat{q}_{l}(0)\sin \omega _{l}t),
\end{eqnarray}%
\\[-2mm]
where $\Delta _{c}=\omega _{c}-\omega _{0}$ is the cavity detuning, $\Delta
_{k}=\omega _{k}-\omega _{0}$ is the detuning of the $k$-th mode of the
environment, and $\Delta _{m}=\omega _{m}+\sum_{l}\omega _{l}\gamma _{l}^{2}$
is the reservoir-induced potential energy shift. The non-Markovian effect is
fully manifested in Eqs.(1), where the non-local time correlation functions
of the environments, i.e., $f_{c}(t)=\sum_{k}g_{k}^{\ast }g_{k}e^{-i\Delta
_{k}t}$ and $f_{m}(t)=\sum_{l}\omega _{l}^{2}\gamma _{l}^{2}\sin \omega
_{l}t $, are included. By introducing the spectral density $J(\omega )$ of
the reservoirs, one can rewrite the time correlation functions as $%
f_{c}(t)=\int \frac{d\omega }{2\pi }J_{c}(\omega )e^{-i(\omega -\omega
_{0})t}$ and $f_{m}(t)=\int \frac{d\omega }{2\pi }J_{m}(\omega )\sin \omega
t $. The terms containing reservoir operators $\hat{q}_{l}(0)$, $\hat{p}%
_{l}(0) $ and $\hat{a}_{k}(0)$ are usually regarded as the noise-input of
the system, which depend on the initial states of the reservoirs.

The integro-differential Heisenberg-Langevin equations Eqs.(1) are
intrinsically nonlinear. Up to now, most experimental realizations of cavity
optomechanics are still in the single-photon weak coupling limit \cite%
{Chan478 89,Teufel475 359,Arcizet444 71,Groblacher460 724}, i.e., $%
10^{-5}\lesssim g_{0}/\omega _{m}\lesssim 10^{-3}$. When the intracavity
photon number $|\langle \hat{a}\rangle |\gg 1$, we can apply the so-called
linearization method \cite{Genes77 033804,Aspelmeyer86 1391}\textbf{. } This
means the relevant quantum operators can be expanded about their respective
mean values: $\hat{O}=\langle \hat{O}\rangle +\delta \hat{O}$, where $\hat{O}%
\equiv (\hat{a},\hat{a}^{\dagger },\hat{q},\hat{p})^{T}$, and the
superscript $T$ represents the transpose operation. Then Eqs.(1) can be
decomposed into two parts. The first is the classical part that describing
the classical phase space orbits of the first moments of operators
\end{subequations}
\begin{eqnarray}
\dot{\alpha} &=&-i\Delta _{c}\alpha +ig_{0}\alpha q-\int_{0}^{t}d\tau
f_{c}(t-\tau )\alpha (\tau )+E,~  \label{classical} \\
\ddot{q} &=&-\omega _{m}\Delta _{m}q+g_{0}\omega _{m}|\alpha |^{2}+\omega
_{m}\int_{0}^{t}d\tau f_{m}(t-\tau )q(\tau ),  \notag
\end{eqnarray}%
where, for simplicity, we have assumed $\langle \hat{q}_{l}(0)\rangle
=\langle $ $\hat{p}_{l}(0)\rangle $ $=\langle \hat{a}_{k}(0)\rangle =0$. In
the single-photon weakly coupling regime, the coupling strength $g_{0}$ is
the smallest parameter in Eqs.(\ref{classical}). We therefore perform the
regular perturbation expansion in ascending powers of the rescaled
dimensionless variable $g_{0}/\omega _{m}$ (for computational convenience
one may set $\omega _{m}=1$, and the other rescaled parameters are in units
of $\omega _{m}$). By substituting the expressions with rescaled $g_{0}$
(i.e., $\alpha =\sum_{n=0}^{\infty }g_{0}^{n}\alpha _{n}$ and $%
q=\sum_{n=0}^{\infty }g_{0}^{n}q_{n}$) into the averaged Langevin equations (%
\ref{classical}), one can give a formal solution up to the first order for
the classical part in the framework of modified Laplace transformation%
\textbf{\ }\cite{Zhang109 170402}
\begin{subequations}
\begin{eqnarray}
\alpha _{0}(t) &=&\alpha (0)\bar{\alpha}_{0}(t)+E\int_{0}^{t}d\tau \bar{%
\alpha}_{0}(\tau ), \\
q_{0}(t) &=&\int_{-\infty }^{\infty }\frac{d\omega }{2\pi }\frac{[i\omega
q(0)-p(0)]e^{-i\omega t}}{\omega ^{2}-\Delta _{m}-K_{m}(\omega )-i\tilde{J}%
(\omega )},~~~~~ \\
\alpha _{1}(t) &=&i\int_{0}^{t}d\tau \bar{\alpha}_{0}(t-\tau )\alpha
_{0}(\tau )q_{0}(\tau ), \\
q_{1}(t) &=&\int_{0}^{t}d\tau \bar{q}_{0}(t-\tau )|\alpha _{0}(\tau )|^{2}
\label{perturbexpan}
\end{eqnarray}%
\\[-2mm]
where $K_{m}(\omega )=\mathcal{P}\int \frac{d\omega ^{\prime }}{2\pi }\frac{%
\omega ^{\prime }J_{m}(\omega ^{\prime })}{\omega ^{2}-\omega ^{\prime 2}}$
and $\tilde{J}(\omega )=\frac{J_{m}(-\omega )-J_{m}(\omega )}{4}$ are the
real and imaginary part of the Laplace transform of the self-energy
correction respectively, and the Green's functions $\bar{\alpha}_{0}$ and $%
\bar{q}_{0}$ obey the Dyson equations (S6) and (S7) with the initial
conditions $\bar{\alpha}_{0}(0)=1$, $\bar{q}_{0}(0)=0$ and $\dot{\bar{q}}%
_{0}(0)=1$ \cite{Zhang109 170402}. Base on Eqs. (3), we can see that the
non-vanishing intracavity field $\alpha _{0}(\infty )$ may induce an
equilibrium position $q_{1}(\infty )$ for the oscillator. This leads to the
effective cavity detuning $\tilde{\Delta}=\Delta _{c}-g_{0}^{2}q_{1}(\infty )
$, which also alters the asymptotic dynamics of the cavity field.
Accordingly, the interplay between the non-Markovian and nonlinear effects
can be described more precisely in this way. Within the parameter space of
our consideration, $\tilde{\Delta}\approx \Delta _{c}$, the validity of the
power series assumption is guaranteed by the numerical simulations (see Sec.
II in the Supplemental Material).

For general bosonic environments, the spectral density should be a
Poisson-type distribution function \cite{Leggett59 1}. We consider that the
spectrum is of the form
\end{subequations}
\begin{equation*}
J_{a}(\omega )=2\pi \eta _{a}\omega (\frac{\omega }{\tilde{\omega}_{a}}%
)^{s_{a}-1}e^{-\frac{\omega }{\tilde{\omega}_{a}}}~~~~(a=c,m),
\end{equation*}%
where $\eta _{a}$ is a dimensionless coupling constant between the system
and the environment, and $\tilde{\omega}_{a}$ is a high-frequency cutoff %
\cite{Leggett59 1,Weiss2008}. The parameter $s_{a}$ classifies the
environment as sub-Ohmic $(0<s_{a}<1)$, Ohmic $(s_{a}=1)$, and super-Ohmic $%
(s_{a}>1)$. Using the modified Laplace transformation, one can give an
analytical solution for the nonequilibrium Green's function
\begin{eqnarray}
\bar{\alpha}_{0}(t) &=&\mathcal{Z}e^{-i\omega _{r}t} \\
&&+\frac{2}{\pi }\int_{-\omega _{0}}^{\infty }d\omega \frac{J_{c}(\omega
+\omega _{0})e^{-i\omega t}}{4[\omega -\Delta _{c}-K_{c}(\omega
)]^{2}+J_{c}^{2}(\omega +\omega _{0})}  \notag
\end{eqnarray}%
with $K_{c}(\omega )=\mathcal{P}\int \frac{d\omega ^{\prime }}{2\pi }\frac{%
J_{c}(\omega ^{\prime })}{\omega -\omega ^{\prime }+\omega _{0}}$. The first
term survives only when\textbf{\ $\omega _{c}+K_{c}(-\omega _{0})\leq 0$},%
\textbf{\ }the\textbf{\ }residue $\mathcal{Z}=[1-K_{c}^{\prime }(\omega
)]^{-1}$, and the pole $\omega _{r}$ is located at the real $z$ axis. This
term \textbf{\ }corresponds to ``localized mode''\ \cite{Zhang109
170402,Cheng91 022328}, which means that the cavity field oscillates with
frequency $\omega _{r}$ and does not decay. It is seem that the photons are
``trapped''\ in the cavity due to the backflow of the non-Markovian
environment and do not diffuse. It is a term that determines the asymptotic
dynamics of the optical field. Physically, this is equivalent \cite{Cheng91
022328} to generate a bound state of the joint cavity-reservoir system,
which can be also determined \cite{Tong81 052330} by solving the energy
eigenstates of the total Hamiltonian, and such bound state is actually a
stationary state with a vanishing decay rate during the time evolution. The
second term corresponds to nonexponential decays. In the long time limit,
the bound state as well as the driving laser give rise to the non-vanishing
intracavity photon numbers with $\alpha _{0}(\infty )=\frac{\omega
_{r}\alpha (0)+iE}{\omega _{r}[1-K_{c}^{\prime }(\omega _{r})]}e^{-i\omega
_{r}t}-\frac{iE}{\Delta _{c}+K_{c}(0)-\frac{i}{2}J_{c}(\omega _{0})}$. On
the other hand, for the mechanical mode, however, due to the discontinuity
of the self-energy correction at the real $z$ axis, $q_{0}(\infty )=0$. We
find that the corresponding mechanical bound state can be formed only if
some sharp cutoff appear in the spectral density (see Supplemental Material
Sec. II).

\begin{figure}[tp]
{\subfigure{\includegraphics[height=4.3cm]{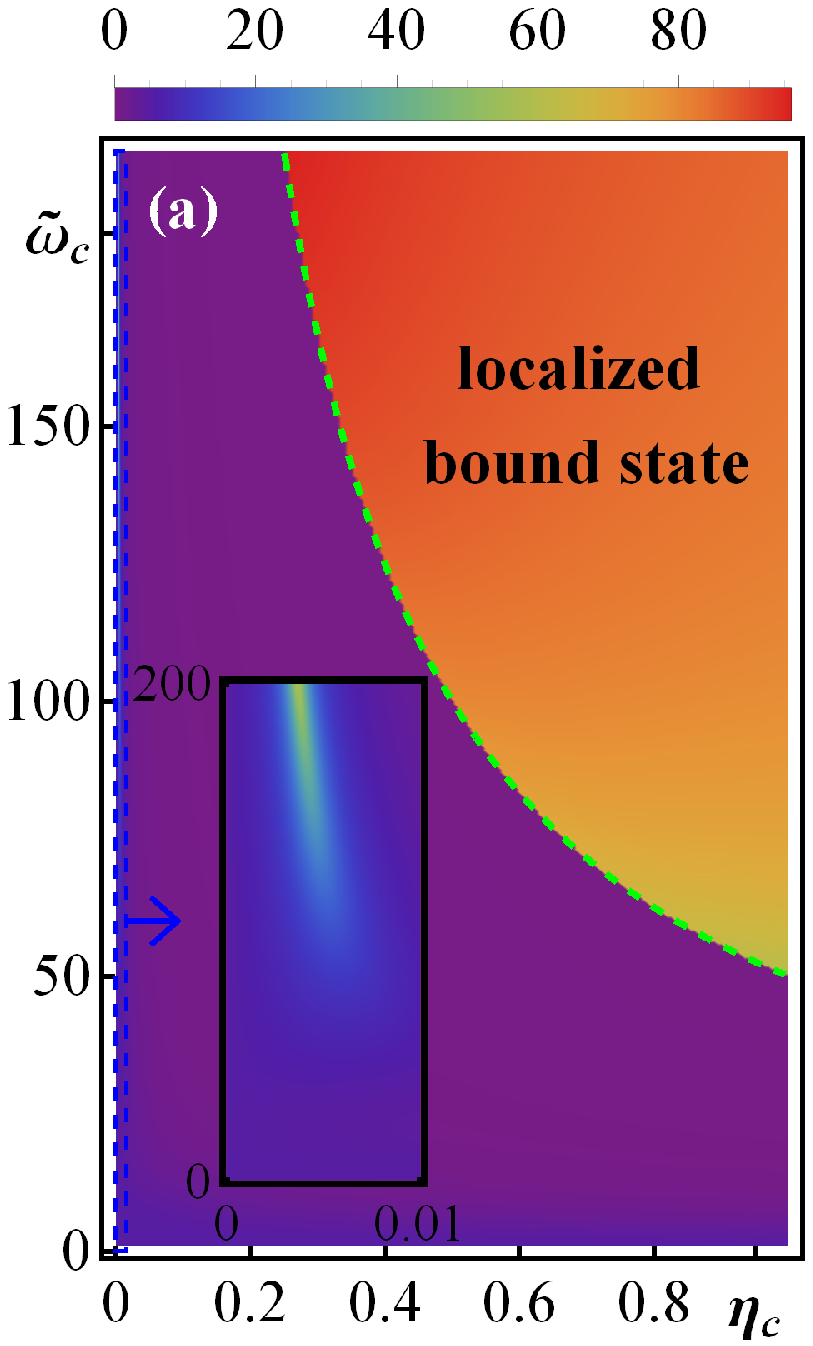}}}\hspace{-1.2ex} {\subfigure{%
\includegraphics[height=4.2cm]{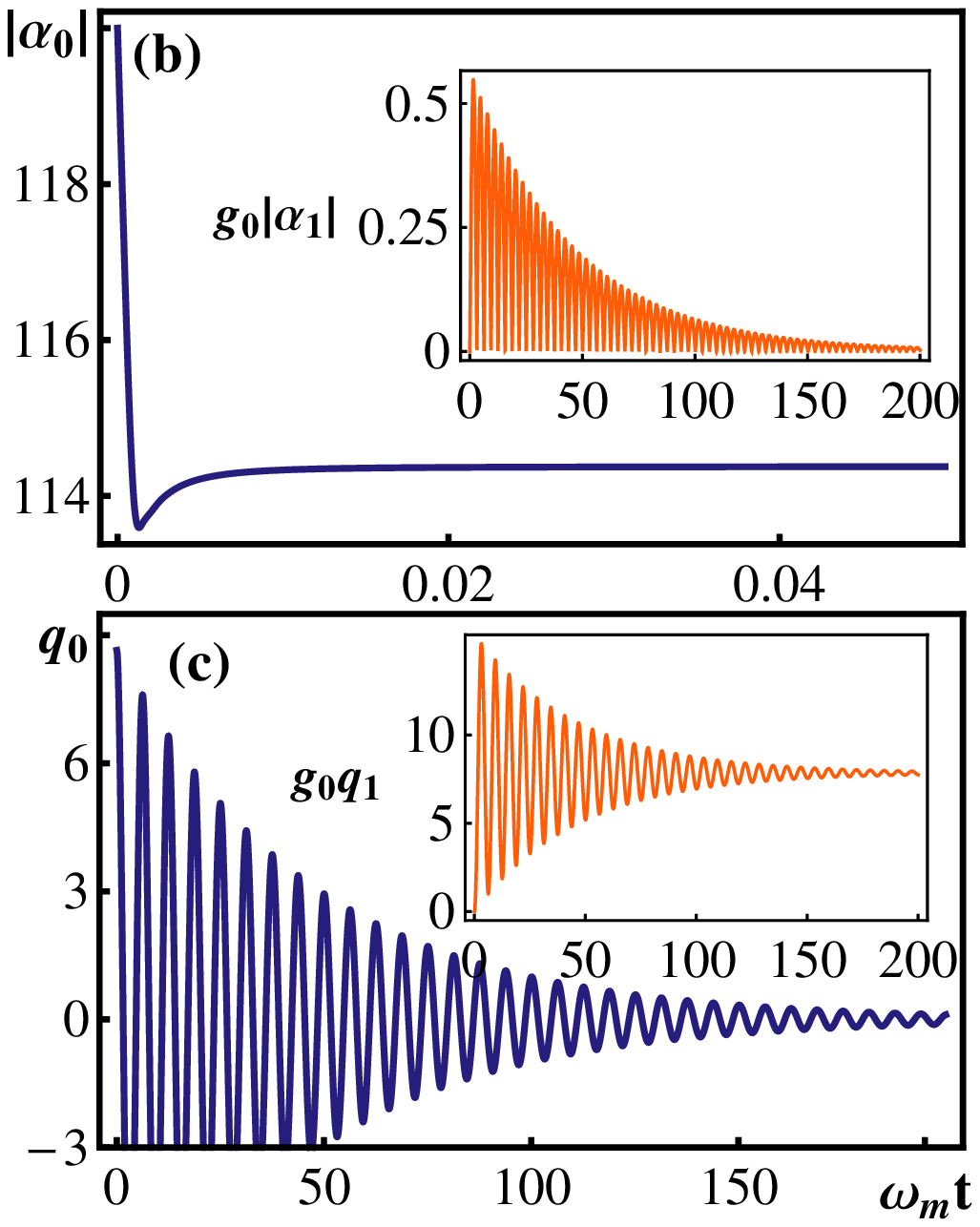}}\hspace{-1.2ex} \subfigure{%
\includegraphics[height=4.2cm]{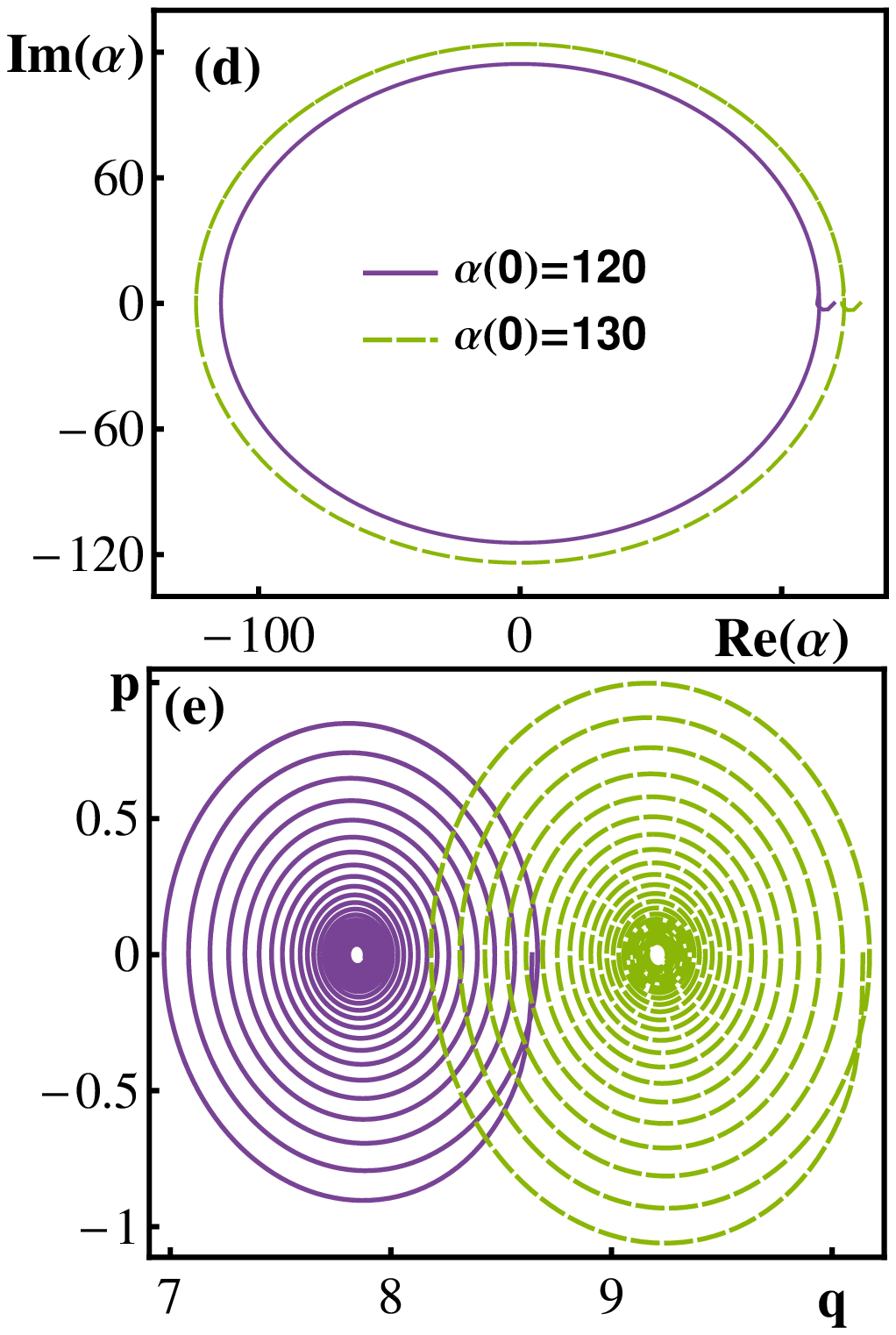}}}
\caption{(Color online) Classical dynamics in an optomechanical system, in
units of $\protect\omega _{m}$. (a) is the density plot of $|\protect\alpha %
_{0}|_{max}(\infty )$, in which we show the region in the parameter space of
coupling $\protect\eta _{c}$ and frequency cutoff $\tilde{\protect\omega}%
_{c} $, where significant bound state exist with $E=10\protect\omega _{m}$
and $\Delta _{c}=2\protect\omega _{m}$. Plot (b-e) show the dynamical
evolution of the classical variables, where $g_{0}=6\times 10^{-4}\protect%
\omega _{m}$, $\ \protect\eta _{m}=0.03$, $\tilde{\protect\omega}_{m}=11%
\protect\omega _{m} $ and $E=0$. The other parameters are $\protect\eta %
_{c}=0.05$, $s_{c}=3$, $s_{m}=1$, $\tilde{\protect\omega}_{c}=11\protect%
\omega _{c}=1100\protect\omega _{m}$, $\protect\alpha (0)=120$, $p(0)=0$,
and we keep $q(0)=\frac{g_{0}}{\protect\omega _{m}}|\protect\alpha (0)|^{2}$%
. }
\label{f1}
\end{figure}

Considering the feasibility, the single photon coupling $g_{0}$
(key parameter) is set close to that of recently performed optomechanical
experiments \cite{Aspelmeyer86 1391}. Fig. \ref{f1}(a) is the density plot
of the maximum value of $|\alpha _{0}|$ in the long-time limit. It maps out
the regions in parameter space where localized bound state occurs. As a
result, a threshold characterizing the transition from weak to strong
non-Markovian regions can be defined, and it is marked by the green-dashed
line, which satisfying $\omega _{c}+K_{c}(-\omega _{0})=0$. In the weakly
non-Markovian region $0<\eta _{c}<0.01$, as shown in the inset of Fig. \ref%
{f1}(a), the red-detuned laser gives rise to a strong stationary amplitude
values, which is determined by $\frac{iE}{\frac{i}{2}J_{c}(\omega
_{0})-\Delta _{c}-K_{c}(0)}$. Figs. \ref{f1}(b) and \ref{f1}(c) show the
dynamic evolution of $|\alpha _{0}|$ and $q_{0}$, where the first order
solutions are shown in the insets. In Fig. \ref{f1}(b) the optical bound
state is formed when $\eta _{c}$ is above the threshold, while for Fig. \ref%
{f1}(c), the coordinate average value of the oscillator is no longer zero,
which reveals that radiation pressure push the oscillator to a new
equilibrium position. The evolution in the phase space is shown in Figs.
2(d) and 2(e). The limit cycles that characterize the properties of the
steady state depends on the initial states, which indicates the
non-Markovian property and reflects the memory effect. The initial
information of the system is maintained.

We now turn to the quantum fluctuation of operators that describes the
actual quantum dynamics, which are deduced as%
\begin{equation}
\delta \dot{\hat{O}}=M(t)\delta \hat{O}-\int_{0}^{t}d\tau F(t-\tau )\delta
\hat{O}(\tau )+\hat{\xi}(t).  \label{fluctuation}
\end{equation}%
By assuming the formal solution $\delta \hat{O}=\mathcal{U}(t)\delta \hat{O}%
(0)+\hat{\mathcal{V}}(t)$, and substituting it into Eq. (\ref{fluctuation}),
we have
\begin{subequations}
\begin{eqnarray}
\dot{\mathcal{U}}(t) &=&M(t)\mathcal{U}(t)-\int_{0}^{t}d\tau F(t-\tau )%
\mathcal{U}(\tau ), \\
\dot{\hat{\mathcal{V}}}(t) &=&M(t)\hat{\mathcal{V}}(t)-\int_{0}^{t}d\tau
F(t-\tau )\hat{\mathcal{V}}(\tau )+\hat{\xi}(t),~~~
\end{eqnarray}%
\\[-2mm]
subjected to the initial conditions $\mathcal{U}(0)=\mathbf{I}$ and $\hat{%
\mathcal{V}}(0)=0$. The $4\times 4$ matrix
\end{subequations}
\begin{equation*}
M(t)=\left(
\begin{array}{cccc}
-i\Delta _{cq}(t) & 0 & ig_{0}\alpha (t) & 0 \\
0 & i\Delta _{cq}(t) & -ig_{0}\alpha ^{\ast }(t) & 0 \\
0 & 0 & 0 & \omega _{m} \\
g_{0}\alpha ^{\ast }(t) & g_{0}\alpha (t) & -\Delta _{m} & 0%
\end{array}%
\right)
\end{equation*}%
describes the linearized optomechanical coupling with non-local
time-dependent classical variables, where $\Delta _{cq}(t)=\Delta
_{c}-g_{0}q(t)$ and $\Delta _{m}=\omega _{m}+\eta _{m}\tilde{\omega}%
_{m}\Gamma (s_{m})$. The matrix
\begin{equation*}
F(t)=\left(
\begin{array}{cccc}
f_{c}(t) & 0 & 0 & 0 \\
0 & f_{c}^{\ast }(t) & 0 & 0 \\
0 & 0 & 0 & 0 \\
0 & 0 & -f_{m}(t) & 0%
\end{array}%
\right)
\end{equation*}%
depicts the time correlation of the optomechanical system in the bosonic
environments. The last term $\hat{\xi}(t)=(-i\sum_{k}g_{k}e^{-i\Delta _{k}t}%
\hat{a}_{k}(0),i\sum_{k}g_{k}^{\ast }e^{i\Delta _{k}t}\hat{a}_{k}^{\dag
}(0),0,\sum_{l}\omega _{l}\gamma _{l}[\hat{p}_{l}(0)\newline
\times \cos \omega _{l}t-\hat{q}_{l}(0)\sin \omega _{l}t])^{T}$ is
interpreted as a noise term \cite{Giovannetti63 023812,Genes77 033804} that
depends on the initial states of the environments. It is easy to obtain the
quadrature operator $\hat{R}\equiv (\delta \hat{x}_{c},\delta \hat{p}%
_{c},\delta \hat{q},\delta \hat{p})^{T}$ through the relation $\hat{R}=%
\mathcal{S}\delta \hat{O}$, where $\mathcal{S}$ is the transformation
matrix. Then the covariance matrix with components $V_{ij}=\langle \hat{R}%
_{i}\hat{R}_{j}+\hat{R}_{j}\hat{R}_{i}\rangle /2$ can be determined by
calculating the time evolution of the second moments of the quadratures
\begin{eqnarray}
\hat{R}\hat{R}^{T} &=&\mathcal{S}\mathcal{U}\mathcal{S}^{-1}\hat{R}(0)\hat{R}%
^{T}(0){\mathcal{S}^{-1}}^{T}\mathcal{U}^{T}\mathcal{S}^{T}+\mathcal{S}\hat{%
\mathcal{V}}\hat{\mathcal{V}}^{T}\mathcal{S}^{T} \\
&&+\mathcal{S}\mathcal{U}\mathcal{S}^{-1}\hat{R}(0)\hat{\mathcal{V}}^{T}%
\mathcal{S}^{T}+\mathcal{S}\hat{\mathcal{V}}\hat{R}^{T}(0){\mathcal{S}^{-1}}%
^{T}\mathcal{U}^{T}\mathcal{S}^{T}.~~~  \notag
\end{eqnarray}%
The first term is the projection of the quadratures on the system's Hilbert
space. The second term characterizes the magnitudes of the input-noise that
satisfies the non-local time correlation relations \cite{Giovannetti63
023812}. The last two terms describe the effect of initial system-reservoir
correlations, which had been identified as an important factor in the
decoherence dynamics \cite{Romero55 4070,Dijkstra104 250401}. For the sake
of simplicity, here we assume as usual the system and the reservoirs are
initially uncorrelated, and the reservoirs are in thermal states. Then the
noise vector $\hat{\xi}(t) $ obeys the non-Markovian self-correlation $%
\langle \hat{\xi}(t)\hat{\xi}(t^{\prime })\rangle =G_{c}(t-t^{\prime
})\oplus G_{m}(t-t^{\prime })$, where $G_{c}$ and $G_{m}$ are $2\times 2$
matrix
\begin{equation*}
G_{c}(\tau )=\left(
\begin{array}{cc}
0 & \tilde{g}_{c}(\tau ) \\
g_{c}(-\tau ) & 0%
\end{array}%
\right) ,~~~G_{m}(\tau )=\left(
\begin{array}{cc}
0 & 0 \\
0 & g_{m}(\tau )%
\end{array}%
\right) .
\end{equation*}%
The thermal correlation functions are defined as $g_{c}(\tau )=\int \frac{%
d\omega }{2\pi }J_{c}(\omega )\frac{e^{-i(\omega -\omega _{0})\tau }}{%
e^{\beta _{c}\omega }-1}$, $\tilde{g}_{c}(\tau )=\int \frac{d\omega }{2\pi }%
J_{c}(\omega )\frac{e^{\beta _{c}\omega -i(\omega -\omega _{0})\tau }}{%
e^{\beta _{c}\omega }-1}$ and $g_{m}(\tau )=\int \frac{d\omega }{4\pi }%
J_{m}(\omega )(\frac{2\cos {\omega \tau }}{e^{\beta _{m}\omega }-1}%
+e^{-i\omega \tau })$, where $\beta =1/k_{B}T$, $k_{B}$ is the Boltzmann
constant and $T$ is the initial temperature of the reservoir.

\textit{Preservation of entanglement} To bring quantum effects to the
macroscopic level, one important way is the creation of entanglement between
the optical mode and the mechanical mode. If the initial states of the
system is Gaussian, then Eq. (\ref{fluctuation}) will preserve the Gaussian
character. The entanglement can therefore be quantified via the logarithmic
negativity \cite{Adesso72 032334} defined as $E_{\mathcal{N}}=\max [0,-\ln
(2s^{-})]$, where $s^{-}$ is the smallest symplectic eigenvalues of the
partially transposed covariance matrix $V^{T_{A}}$.

\begin{figure}[tbp]
{\subfigure{\includegraphics[width=4.6cm]{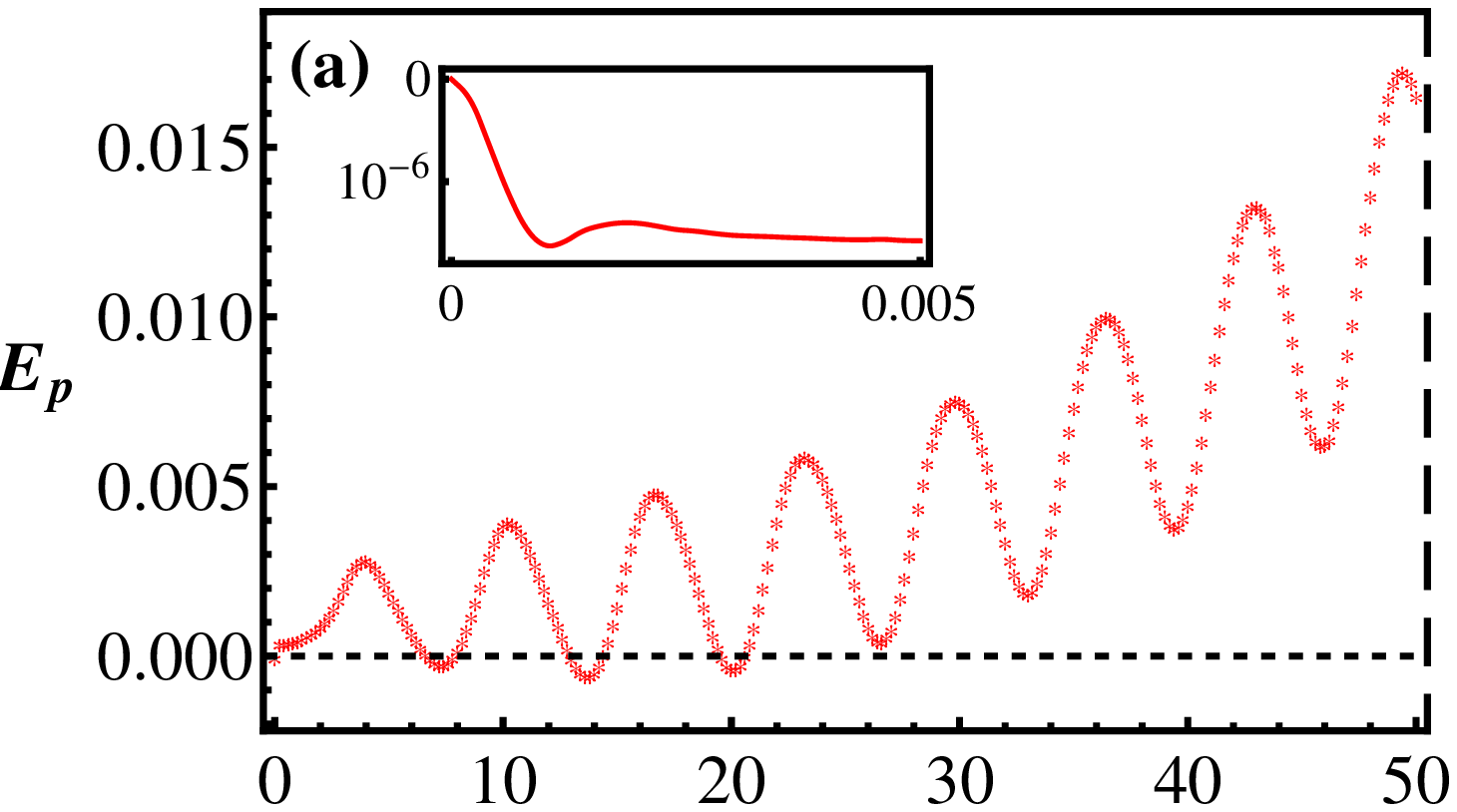}} \subfigure{%
\includegraphics[width=3.7cm]{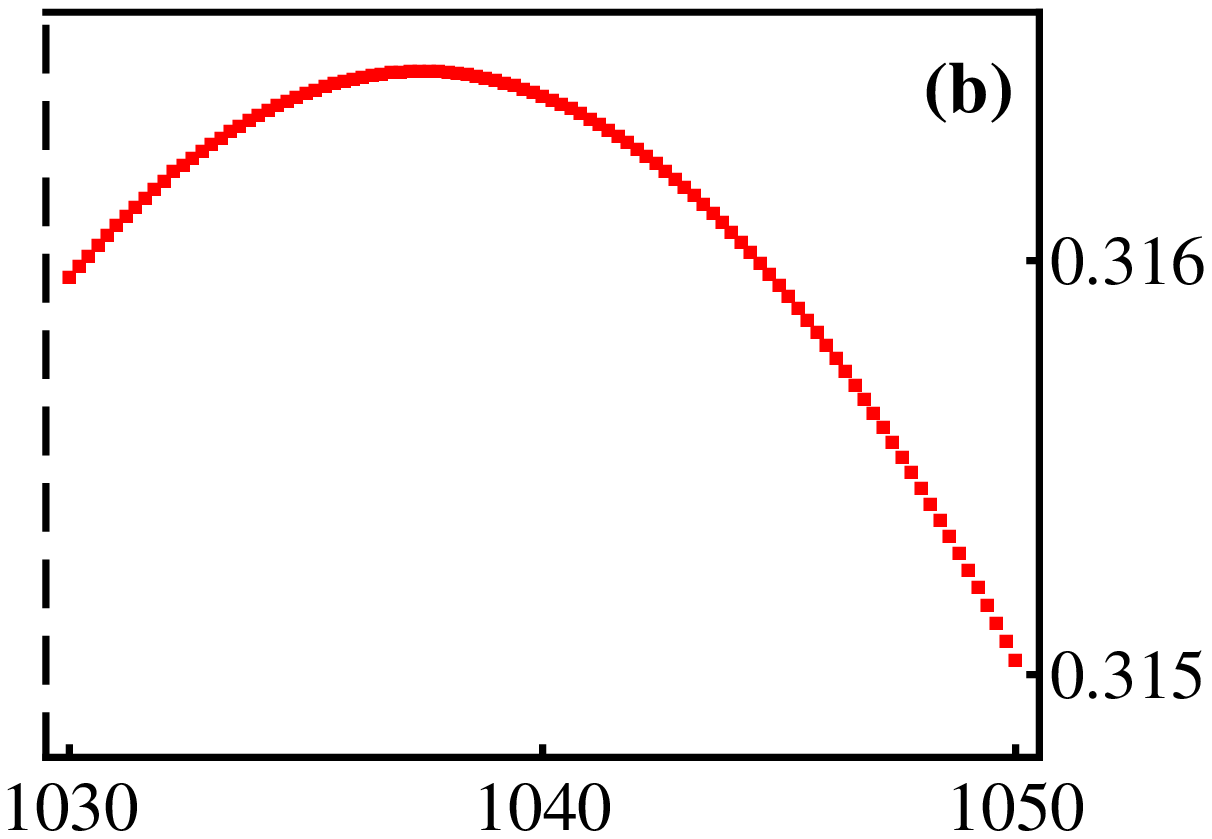}}\vspace{-2.ex}} {\subfigure{%
\includegraphics[width=4.6cm]{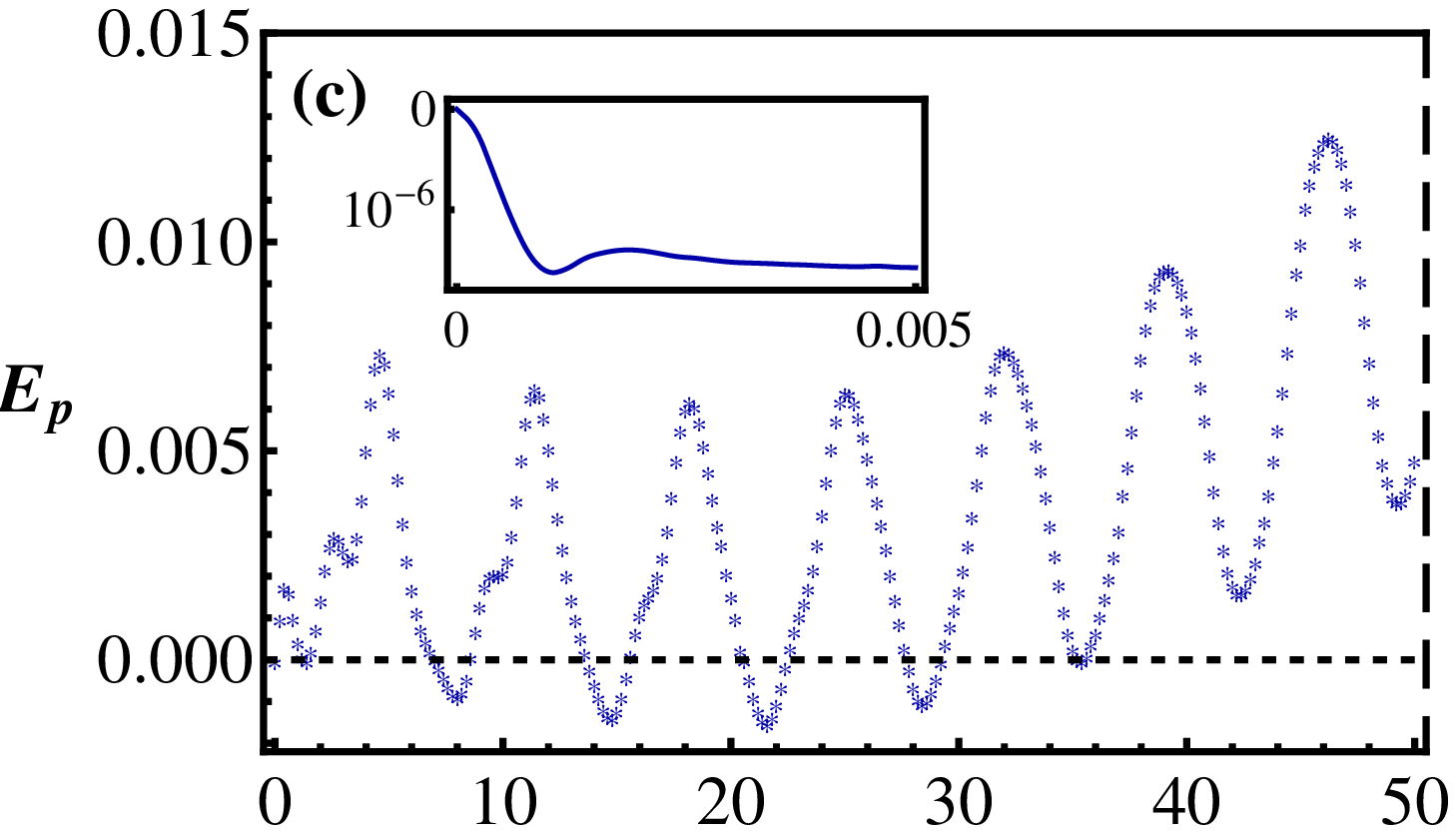}} \subfigure{%
\includegraphics[width=3.7cm]{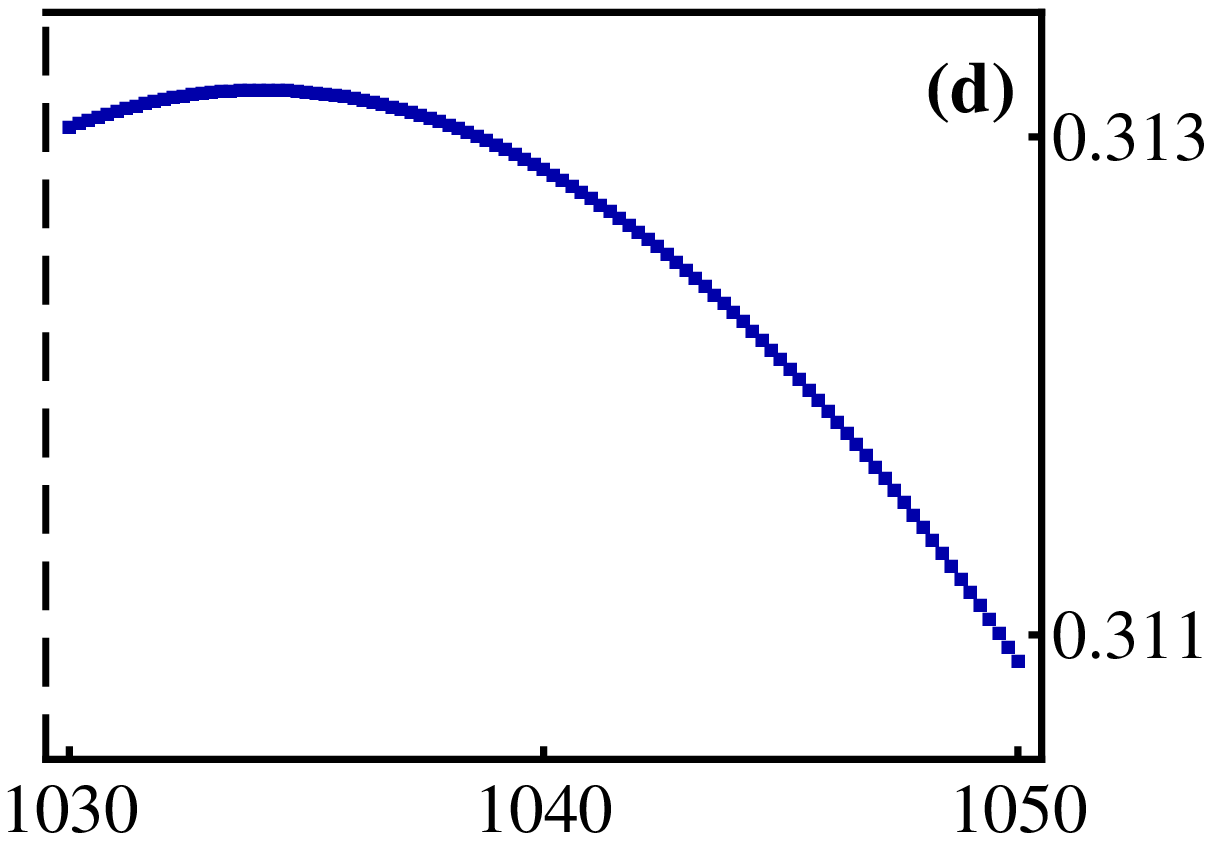}} \vspace{-2ex}~~~~~~~~~~~ $\omega _{m}t$%
}~~~~~~~~~~~ \hspace{-5.6ex}
\caption{(Color online) Time evolution of pseudoentanglement $E_{p}$ in
Ohmic environment. In (a) and (b), we keep $s_{m}=1$, while in (c) and (d), $%
s_{m}=3$, $\protect\eta _{m}=0.8$ and $\tilde{\protect\omega}_{m}=5\protect%
\omega_{m}$. The other parameters are the same as Fig. 2(b). The dynamical
evolution of entanglement are shown in three regions, the initial stage of
evolution $\protect\omega _{m}t<5\times10^{-3}$ shown in the insets of (a)
and (c) , the short-time scale $\protect\omega _{m}t<50$ corresponding to
(a) and (c), the long-time scale $\protect\omega _{m}t$ around $1040$ for
(b) and (d). The regions $E_{p}<0$ correspond to nonphysical results.}
\label{ep}
\end{figure}

The non-Markovian effect generally lead to the nonequilibrium dynamics,
therefore, we explore the optomechanical entanglement in transient regime.
In order to show clearly the evolutionary path of the entanglement, we use
the so-called pseudoentanglement measure \cite{Wang112 110406} defined as $%
E_{p}=-\ln (2s^{-})$, so that the logarithmic negativity of entanglement
measure $E_{\mathcal{N}}=\max [0,E_{p}]$. For simplicity, we assume the bath
of the cavity initially in vacuum and the cavity initially in coherent state
with $\alpha (0)=120$. The results are plotted in Fig. 3, in which $\beta
_{m}=2.5\times 10^{-2}\omega _{m}^{-1}$ (the corresponding thermal
excitation number are $n_{m}\approx 40$). In order to show clearly the
phenomenon of entanglement preservation by non-Markovianity, we shutoff the
classical pumping field all the time by setting $E=0$. We see in Fig. 3 that
the shape of the Ohmic spectrum characterized by $s_{m}$, has effect on the
entanglement dynamics in the short-time and long-time scale, but has
negligible impact on the initial stage of evolution (see the insets). This
is because the time scale of the mechanical oscillator as well as its bath
is much larger than that of the bath of the cavity, which means the
non-Markovian memory effects induced by the bath of the oscillator can be
ignored completely when $t\ll 1/d$, where $d$ is the bandwidth of the
oscillator's bath. Although the sudden death and rebirth of entanglement is
also observed, it differ with the case of Markovian environment \cite{Yu323
598} where the photons would rapidly dissipate to the memoryless
environment. Here, due to the present of the bound state, the entanglement
can be produced and can be preserved in non-Markovian environment after long
time even without external drive. This provides a way to decoherence control
of optomechanical systems, in which, a controllable quantum environment indeed 
have the ability to protect the quantum correlation of the internal system.

\textit{Conclusion} We have put forward a scheme to preserve the
entanglement of optomechanical system in non-Markovian environment. An
analytical approach for describing non-Markovian memory effects that impact
on the decoherence dynamics of an optomechanical system is presented. The
exact Heisenberg-Langevin equations are derived, and the perturbation
solution is given in the weak single-photon coupling regime. Employing the
analytical solution, we have shown that, the system dynamics change
dramatically when the cavity-bath coupling strength crosses a certain
threshold, which corresponds to dissipationless non-Markovian dynamics. The
interplay between non-Markovian and nonlinear effects can be also explained
though the perturbative method. As a quantum device which may subjected to
dissipative and decoherence effects, however, our results show that the
surroundings of such a simple setting can protect the quantum entanglement,
rather than destroy it even in the long-time scales, which means that
engineered quantum noise can be used in robust quantum state generation. Our
research provides a new approach to explore non-Markovian dynamics for the
cavity optomechanical systems.


\textit{Acknowledgment} We would like to thank B.-L. Hu, P.-Y. Lo, H.-N.
Xiong and W.-L. Li for helpful discussions. This work is supported by the
NSF of China under Grant No. 11474044.



\begin{thebibliography}{99}
\bibitem{Breuer2002} H.-P. Breuer and F. Petruccione, \textit{The Theory of
Open Quantum Systems} (Oxford University Press, Oxford, UK, 2002).


\bibitem{Weiss2008} U. Weiss, \textit{Quantum Dissipative Systems}, 3rd ed.
(World Scientific Press, Singapore, 2008).

\bibitem{DiVincenzo393} D. P. DiVincenzo, Nature 393, 113 (1998).

\bibitem{Knill409 46} E. Knill, R. Laflamme, and G. J. Milburn, Nature 409,
46 (2001).

\bibitem{cirac} F. Verstraete, M. M. Wolf, and J. I. Cirac, Nature Phys. 5,
633 (2009).

\bibitem{Eisert} M. J. Kastoryano, M. M. Wolf, and J. Eisert, Phys. Rev.
Lett. 110, 110501 (2013).

\bibitem{Chru104 070406} D. Chru\'{s}ci\'{n}ski and A. Kossakowski, Phys.
Rev. Lett. 104, 070406 (2010).

\bibitem{Xu104 100502} J.-S. Xu, et al., Phys. Rev. Lett. 104, 100502 (2010).

\bibitem{Liu7 931} B.-H. Liu, et al., Nat. Phys. 7, 931 (2011).

\bibitem{Chin109 233601} A. W. Chin, S. F. Huelga, and M. B. Plenio, Phys.
Rev. Lett. 109, 233601 (2012).

\bibitem{Deffner111 010402} S. Deffner and E. Lutz, Phys. Rev. Lett. 111,
010402 (2013).

\bibitem{Rivas105 050403} \'{A}. Rivas, S. F. Huelga, and M. B. Plenio,
Phys. Rev. Lett. 105, 050403 (2010).

\bibitem{Breuer103 210401} H.-P. Breuer, E.-M. Laine, and J. Piilo, Phys.
Rev. Lett. 103, 210401 (2009).

\bibitem{Vasile84 052118} R. Vasile, S. Maniscalco, M. G. A. Paris, H.-P.
Breuer, and J. Piilo, Phys. Rev. A 84, 052118 (2011).

\bibitem{Lorenzo88 020102} S. Lorenzo, F. Plastina, and M. Paternostro,
Phys. Rev. A 88, 020102 (2013).

\bibitem{Chru112 120404} D. Chru\'{s}ci\'{n}ski and S. Maniscalco, Phys.
Rev. Lett. 112, 120404 (2014).

\bibitem{Vitali98 030405} D. Vitali et al., Phys. Rev. Lett. 98, 030405
(2007).

\bibitem{Kippenberg321 1172} T. J. Kippenberg and K. J. Vahala, Science 321,
1172 (2008).

\bibitem{Groblacher460 724} S. Gr\"{o}blacher, K. Hammerer, M. R. Vanner,
and M. Aspelmeyer, Nature (London) 460, 724 (2009).

\bibitem{Connell464 697} A. D. O'Connell et al., Nature (London) 464, 697
(2010).

\bibitem{Thompson452 900} J. D. Thompson et al., Nature 452, 900 (2008).

\bibitem{Giovannetti63 023812} V. Giovannetti and D. Vitali, Phys. Rev. A
63, 023812 (2001).

\bibitem{Genes77 033804} C. Genes, D. Vitali, P. Tombesi, S. Gigan, and M.
Aspelmeyer, Phys. Rev. A 77, 033804 (2008).

\bibitem{Agarwal81 041803} G. S. Agarwal and S. Huang, Phys. Rev. A 81,
041803 (2010).

\bibitem{Rabl107 1} P. Rabl, Phys. Rev. Lett. 107, 063601 (2011). 

\bibitem{Bayindir84 2140} M. Bayindir, B. Temelkuran, and E. Ozbay, Phys.
Rev. Lett. 84, 2140 (2000).

\bibitem{Hartmann2 849} M. J. Hartmann, F. G. S. Brandao, and M. Plenio,
Nature Phys. 2, 849 (2006).

\bibitem{Groblacher6 7606} S. Gr\"{o}blacher, A. Trubarov, N. Prigge, G. D.
Cole, M. Aspelmeyer and J. Eisert, Nature Commun. 6, 7606 (2015).

\bibitem{Wu18 18407} M. H. Wu, C. U. Lei, W. M. Zhang, and H. N. Xiong, Opt.
Express 18, 18407 (2010).


\bibitem{Ford37 4419} G. W. Ford, J. T. Lewis, and R. F. O'Connell, Phys.
Rev. A 37, 4419 (1988).

\bibitem{Chan478 89} J. Chan, T. P. M. Alegre, A. H. Safavi-Naeini, J. T.
Hill, A. Krause, S. Gr\"{o}blacher, M. Aspelmeyer, and O. Painter, Nature
(London) 478, 89 (2011).

\bibitem{Teufel475 359} J. D. Teufel et al.
Nature (London) 475, 359 (2011).


\bibitem{Arcizet444 71} O. Arcizet, P.-F. Cohadon, T. Briant, M. Pinard, and
A. Heidmann, Nature (London) 444, 71 (2006).

\bibitem{Aspelmeyer86 1391} M. Aspelmeyer, T. J. Kippenberg, and F.
Marquardt, Rev. Mod. Phys. 86, 1391 (2014).

\bibitem{Zhang109 170402} W. M. Zhang, P. Y. Lo, H. N. Xiong, Matisse
Wei-Yuan Tu , and F. Nori, Phys. Rev. Lett. 109, 170402 (2012).

\bibitem{Leggett59 1} A. J. Leggett, S. Chakravarty, A. T. Dorsey, M. P. A.
Fisher, A. Garg and W. Zwerger, Rev. Mod. Phys. 59, 1 (1987).

\bibitem{Cheng91 022328} J. Cheng, W.-Z. Zhang, Y. Han, and L. Zhou, Phys.
Rev. A 91, 022328 (2015).

\bibitem{Tong81 052330} Q. J. Tong, J. H. An, H. G. Luo, and C. H. Oh, Phys.
Rev. A 81, 052330 (2010).




\bibitem{Romero55 4070} L. D. Romero and J. P. Paz, Phys. Rev. A 55, 4070
(1997).

\bibitem{Dijkstra104 250401} A. G. Dijkstra and Y. Tanimura, Phys. Rev.
Lett. 104, 250401 (2010).




\bibitem{Adesso72 032334} G. Adesso and F. Illuminati Phys. Rev. A 72,
032334 (2005).

\bibitem{Wang112 110406} G. Wang, L. Huang, Y.-C. Lai, and C. Grebogi, Phys.
Rev. Lett. 112, 110406 (2014).

\bibitem{Yu323 598} T. Yu and J. H. Eberly, Science 323, 598 (2009).
\end{thebibliography}
\end{document}